\documentclass[letterpaper, 10 pt, conference]{ieeeconf}  % Comment this line out if you need a4paper

\IEEEoverridecommandlockouts                              % This command is only needed if 
\usepackage{mathptmx} % assumes new font selection scheme installed
\usepackage{amsmath} % assumes amsmath package installed
\usepackage{amssymb}  % assumes amsmath package installed
\usepackage{graphicx}  % assumes amsmath package installed
\usepackage{tabularx}
\usepackage{booktabs}
\usepackage{caption}
\usepackage{subcaption}
\usepackage{multirow}
\usepackage{multicol}
\title{\LARGE \bf
Trust Recognition in Human-Robot Cooperation Using EEG
}
\author{Caiyue Xu, Changming Zhang, Yanmin Zhou, Zhipeng Wang, Ping Lu, Bin He, \emph{Senior Member, IEEE}\\% <-this % stops a space
\thanks{This work was supported in part by the National Key Research and Development Program of China (No. 2020AAA0108905), in part by the National Natural Science Foundation of China (No. U2013602, 62088101, 51975415, 61825303), in part by the Science and Technology Commission of Shanghai Municipality (No. 2021SHZDZX0100, 22ZR1467100), and the Fundamental Research Funds for the Central Universities. (\emph{Corresponding author: Yanmin Zhou, email: yanmin.zhou@tongji.edu.cn}.)}% <-this % stops a space
\thanks{The authors are all with the Department of Control Science and Engineering, College of Electronics and Information Engineering, Tongji University, Shanghai 201804, China, with National Key Laboratory of Autonomous Intelligent Unmanned Systems, Shanghai 201210, China, and Frontiers Science Center for Intelligent Autonomous Systems, Shanghai 201210,  China.}%
}

\begin{document}

\maketitle
\thispagestyle{empty}
\pagestyle{empty}

%%%%%%%%%%%%%%%%%%%%%%%%%%%%%%%%%%%%%%%%%%%%%%%%%%%%%%%%%%%%%%%%%%%%%%%%%%%%%%%%
\begin{abstract}
Collaboration between humans and robots is becoming increasingly crucial in our daily life. In order to accomplish efficient cooperation, trust recognition is vital, empowering robots to predict human behaviors and make trust-aware decisions. Consequently, there is an urgent need for a generalized approach to recognize human-robot trust. This study addresses this need by introducing an EEG-based method for trust recognition during human-robot cooperation. A human-robot cooperation game scenario is used to stimulate various human trust levels when working with robots. To enhance recognition performance, the study proposes an EEG Vision Transformer model coupled with a 3-D spatial representation to capture the spatial information of EEG, taking into account the topological relationship among electrodes. To validate this approach, a public EEG-based human trust dataset called EEGTrust is constructed. Experimental results indicate the effectiveness of the proposed approach, achieving an accuracy of 74.99\% in slice-wise cross-validation and 62.00\% in trial-wise cross-validation. This outperforms baseline models in both recognition accuracy and generalization. Furthermore, an ablation study demonstrates a significant improvement in trust recognition performance of the spatial representation. The source code and EEGTrust dataset are available at https://github.com/CaiyueXu/EEGTrust.
\end{abstract}
%%%%%%%%%%%%%%%%%%%%%%%%%%%%%%%%%%%%%%%%%%%%%%%%%%%%%%%%%%%%%%%%%%%%%%%%%%%%%%%%
\section{Introduction}
With the advancement of artificial intelligence and robotics, machines are evolving into autonomous systems. These systems are gradually taking over responsibilities from humans in various aspects of daily life, such as household robots and self-driving vehicles. In such dynamic and unstructured scenarios, effective cooperation between robots and humans becomes essential. The cooperation combines the robot's capabilities for comprehensive planning and rapid execution with the cognitive abilities and adaptability of humans. Trust is of critical importance for the effectiveness of the cooperation. It significantly influences human decision-making and the extent to which they rely on autonomous systems \cite{Yuan L}. For efficient and seamless cooperation with humans, autonomous systems should be able to recognize human trust. This recognition allows the systems to predict human behaviors and make trust-aware autonomous decisions \cite{Akash K1}. For instance, they may depend on the perceived level of trust to switch between autonomous and manual modes \cite{Xie Y, Saeidi H}, or adapt their strategies of choosing between high-risk or conservative approaches \cite{Sadrfaridpour B, Xu A, Chen M}.

It is well established that recognizing human trust in machines is a challenging task due to the multidimensional, dynamic, and uncertain nature of trust dynamics \cite{Malle B F}. Researchers have developed various approaches to identify and measure human-robot trust, with subjective reporting being one of the most commonly used methods. This approach involves designing questionnaires with single or multi-question Likert scales to assess the human trust levels \cite{Malle B F, Jian J Y, Merritt S M}. Representative examples of subjective reporting scales include Muir's 4-item Scale and Jian's 12-item Scale \cite{Jian J Y}. Although subjective reporting can intuitively and conveniently detect human trust, it has its limitations. For one, it may interrupt the human-robot interaction process and cannot be used for real-time measurement. Additionally, there can be discrepancies between the reported trust levels and the actual trust behaviors of humans, as Chen's study revealed \cite{J. Y. C. Chen}. Alternatively, human trust can be inferred by observing their behaviors and decisions, guided by the principle of rationality \cite{Baker}. Examples of behavioral measures include physical movement, active boundary entries, and compliance with suggestions \cite{Freedy A, Miller D, Wright T J, Lu Y}. These methods provide real-time and objective measures for human trust. Nevertheless, these methods often have limited applicability to specific scenarios and are challenging to generalize across different tasks. Given these challenges and limitations, there is an emerging need for a generalized and real-time method for recognizing human-robot trust.

Brain activities have been proven correlated to human trust  \cite{Aimone, Baumgartner, Riedl}, including trust in robots \cite{Oh S, Wang M}. 
EEG is a promising solution for real-time recognition of human-robot trust due to its ability to measure the electrical activity in the cerebral cortex with high time resolution. However, there have been relatively few studies in this domain \cite{Akash K, Choo, Ajenaghughrure}. Akash et al. \cite{Akash K} first made an attempt to measure human-robot trust using EEG and Galvanic skin response (GSR), achieving 73\% accuracy. Choo et al. \cite{Choo} proposed a framework using the convolutional neural network (CNN) classifier that achieved an accuracy of 94.3\% for estimating multilevel trust. Ajenaghughrure et al. \cite{Ajenaghughrure} introduced five traditional machine learning-based ensemble trust classifier models and achieved 60\% accuracy. These studies have demonstrated EEG is effective in measuring human-robot trust in real-time and is promising for application in various domains. Nevertheless, we have noticed that existing studies have mainly focused on trust when humans are monitoring the robot's performance. Little attention has been paid to trust within the context of human-robot cooperation scenarios. In addition, these studies have generally treated the EEG data from various channels as sequential data, ignoring their spatial relationships. These spatial relationships have been shown quite related to brain activities \cite{Bassett}, with potential enhancement for trust recognition.

In this study, we aim to achieve measurement of human trust levels during human-robot cooperation using EEG. Rather than simply observing robots, we employ a human-robot cooperation game as a stimulus to induce varying levels of trust. To the best of our knowledge, human-robot cooperation has not been investigated as the stimulus within the context of EEG-based trust recognition before. Regarding the EEG-based trust recognition model, we propose a novel model to capture the spatial information of EEG data, which corresponds to brain activity. Unlike previous studies that treat EEG signals as sequential data, we utilize a 3-D spatial representation for EEG to encode topological relationships among EEG electrodes. With the spatial representation encoding EEG as an image, we adapt the Vision Transformer to introduce spatial attention mechanisms to effectively capture the spatial features.
Our main contributions are as follows:
\begin{itemize}
    \item [(1)] We deploy an experimental design for EEG-based trust recognition during collaborative interactions between humans and robots. This design employs a human-robot cooperation game scenario as the stimulus. Statistical differences demonstrate that our experiment effectively induces varying levels of human trust when cooperating with robots.
    \item[(2)] We propose an EEG Vision Transformer model to enhance EEG-based trust recognition. This model utilizes 3-D spatial representations and spatial attention mechanisms to capture the spatial features of EEG data, ultimately improving the performance of EEG-based human trust recognition.
    \item [(3)] To validate the proposed method for recognizing human trust using EEG, we have constructed a public dataset. Our method achieves 74.99\% and 62.00\% accuracy in the slice-wise and trial-wise cross-validation, demonstrating its superiority over baseline models. Additionally, the ablation study further demonstrates that our model successfully captures spatial information and enhances model performance.
\end{itemize}

% Our proposal involves using 3-D spatial representations of EEG signals and adapting the Vision Transformer to incorporate spatial attention mechanisms. This approach takes into account the topological relationship among EEG electrodes and captures the necessary topological information of the brain for recognizing human trust.

\section{Material and Dataset}
\subsection{Subjective Experiment}
\begin{figure}
    \centering
    \includegraphics[width=1.0\linewidth]{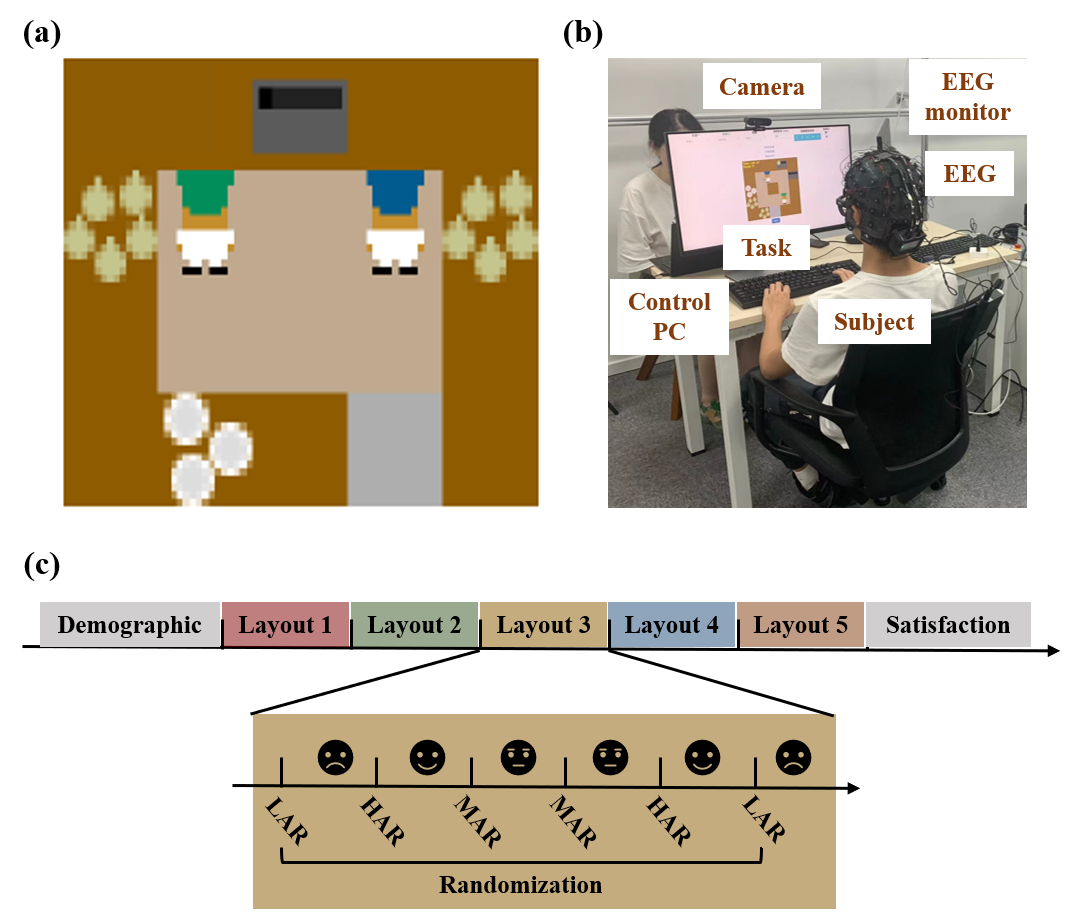}
    \caption{Experimental tasks, setup and protocol for EEGTrust. (a) Overcooked-AI game paradigm. (b) Experimental setup. (c) Experimental protocol.}
    \label{fig1}
\end{figure}
\subsubsection{Participants}
Sixteen undergraduates from Tongji University were recruited in this experiment (2 females, age 24.6 $\pm$ 4.2). All participants were right-handed and had normal or corrected-to-normal vision and hearing. None of the participants had any history of neurological disorders. The study was approved by the institutional review board of Tongji University. All participants signed a consent form before the study. Each participant received \$26 USD as compensation after completing the study.

\subsubsection{Overcooked-AI Game}
The experiment was developed from the Overcooked-AI game, a benchmark environment for fully cooperative human-AI task performance \cite{Carroll}. The game, as depicted in Fig. \ref{fig1}-(a), featured human and autonomous agents that had been pre-trained with reinforcement learning to control one chef each in the kitchen. Their objective was to cook and serve onion soup by placing three onions in a pot, leaving them to cook for twenty seconds, putting the cooked soup in a dish, and serving it, earning both players a reward of 20 scores. To achieve a high reward in the game, the human and AI players should split up the task of delivering onion soups on the fly and cooperate effectively. This required them to perform tasks such as passing items to each other and dividing tasks while also predicting each other's trajectories and avoiding getting in each other's way within a limited shared working space.

\subsubsection{Trust Stimulation and Collection}
Previous studies showed that human trust in a robot is greatly influenced by the robot's ability and the complexity of the task \cite{Hancock}. In this experiment, we used these two factors as independent variables to stimulate different levels of trust. In terms of robot ability, we pre-trained robots with different ability levels using reinforcement learning: high-ability robot (HAR), medium-ability robot (MAR), and low-ability robot (LAR). The abilities included task execution capability, human intent understanding, and cooperation with humans. We also used five different layouts with varying levels of difficulty to test the task complexity. Layouts 1 to 3 were relatively simple, while 4 and 5 were more complex. To collect human trust in the robot, we designed a questionnaire modified from Xu \& Dudek \cite{Xu A}. The questionnaire contained three questions. All questions were designed as 5-point Likert scales. The first two questions investigated participants' subjective feelings regarding the robot's ability to perform the game task and its ability to cooperate with humans. The last question assessed the level of human trust.

\subsubsection{Experiment Protocol}
The experimental setup and protocol are depicted in Fig. \ref{fig1}-(b) and (c). 
We began by briefly explaining the Overcooked-AI game and the task to the participants. Subsequently, participants were asked to complete a survey that included demographic information, EEG-related questionnaires, and Overcooked game familiarity. Following the completion of consent forms and two trial runs for familiarization, the main experiment commenced. In total, there were five layouts and three ability levels of robots. Participants played with each robot twice within each layout. The layouts were presented in a fixed order from one to five. The assignment of robots in each layout was randomized to ensure variability and prevent bias in the experiment.
Consequently, each participant had a total of $5\times3\times2$ trials. Each trial lasted for 60 seconds. Following each trial, participants were required to complete a post-trial questionnaire to provide trust levels. It is worth noting that participants were reminded not to move as much as possible during the experiment and were given 10-second breaks between trials. After the experiment, participants were invited to finish a final questionnaire about their overall satisfaction with the game.

\begin{figure*}
    \centering
    \includegraphics[width=0.8\linewidth]{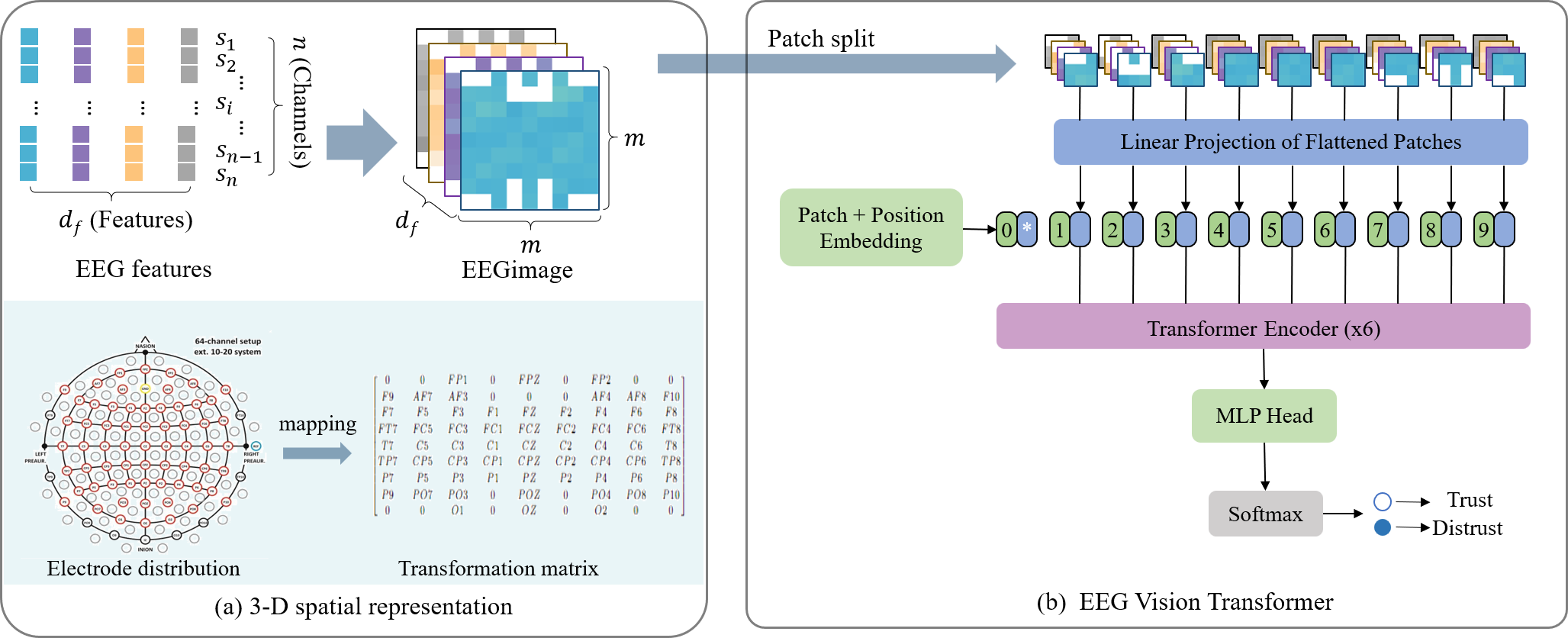}
    \caption{The architecture of our proposed EEG Vision Transformer for trust recognition: (a) The 3-D spatial representation of EEG data. (b) The EEG Vision Transfomer model architecture.}
    \label{fig2}
\end{figure*}

\subsection{Dataset Construction}
\subsubsection{Data Acquisition and Pre-processing}
During the experiment, EEG data were collected using an EEG cap (G.TEC, Inc.) that had 64 active electrodes. The electrodes were organized according to the 10–20 system. The sample rate was set as 250 Hz. After the experiment, the EEG was pre-processed. 
Firstly, the EEG data were re-referenced with the Common Average Reference. Next, EEG data were bandpass filtered between 0.5 to 60 Hz and notch-filtered at 50 Hz to remove the power frequency interference. EOG and EMG artifacts were removed using independent component analysis (ICA) \cite{Hyvärinen}. All of these pre-processing steps were performed using EEGLAB \cite{Delorme} and MATLAB (MathWorks).

\subsubsection{Dataset Description}
EEG data from each trial were split into non-overlapping slices of one second, known as cropped experiments. The reason for doing the cropped experiments was that the predictions of shorter slices were preferred than the trial-wise predictions in real-time recognition \cite{Schirrmeister}.
% This dataset involved 16 subjects, each with 30 trials. Each trial was divided into 60 slices, resulting in 1800 samples per subject. Each sample consisted of 250 (Hz) x 64 (channels) of EEG data. 
The trust level as labels relied on subjective trust scores after each trial. These scores were weighted averaging the scores of the three questions and normalized to trust and distrust. The threshold was adjusted according to each participant's scoring distribution to ensure a balanced dataset.

\section{Trust Vision Transformer Model}
\subsection{Feature Extraction}
In this study, we utilize the differential entropy (DE) as the key feature for analyzing EEG data. DE is an extension of Shannon entropy, specifically designed to quantify the complexity of a continuous random variable. EEG data typically exhibits a higher level of low-frequency energy than high-frequency energy. Thus, DE has a well-balanced ability to distinguish EEG patterns between these low and high-frequency energies. DE was first introduced for EEG-based emotion recognition by Duan et al. \cite{R.-N. Duan}. Building upon their work, we employed a 512-point short-term Fourier transform with non-overlapped Hanning of 1s to extract DE features.
The original calculation formula of DE is defined as follows:
\begin{equation}
    h(X)=-\int_{X}^{} f(x)log(f(x))dx,
\end{equation}
where $X$ is a random variable, $h(X)$ represents the differential entropy of the continuous random variable $X$, and $f(x)$ is the probability density function of $X$. If the random variable $X$ follows a Gaussian distribution $N(\mu,\delta^2)$, the calculation of differential entropy can be simplified as follows:
\begin{equation}
\begin{aligned}
    h(X) &=-\int_{-\propto }^{+\propto } \frac{1}{\sqrt{2\pi \sigma ^{2} } }e^{\frac{(x-\mu)^2}{2\sigma^2} }log\frac{1}{\sqrt{2\pi\sigma^2} }e^{\frac{(x-\mu)^2}{2\sigma^2} }dx \\
     & =\frac{1}{2}log(2\pi e\sigma ^2).
\end{aligned}  
\end{equation}
In the context of EEG analysis, EEG data is typically decomposed into five frequency bands, $\delta$(1-3Hz), $\theta$(4-7Hz), $\alpha$(8-13Hz), $\beta$(14-30Hz), and $\gamma(31-50Hz)$. We extract DE features from EEG data for the last four frequency bands ($\theta, \alpha, \beta, \gamma$). For a specific frequency band $i$, the DE can be defined as
\begin{equation}
    h_i(x)=\frac{1}{2}log(2\pi e\sigma_i ^2).  
\end{equation}

\subsection{3-D Spatial Representation}
An EEG acquisition system involves multiple electrodes that cover the spatial extent of the cerebral cortex. Each electrode has neighbors, and these neighboring electrodes collectively measure the EEG from the underlying brain regions. 
% Traditionally, studies often treat EEG data from multiple channels as a sequence, which ignores the spatial relationships among electrodes.
In this study, we propose a spatial representation to introduce spatial information by mapping the EEG sequence to a mesh grid based on the physical distribution of the electrodes. Specifically, after feature extraction, the EEG data is represented as a sequence ${[s_1, s_2,\cdots, s_n]}^T\in\mathbb R^{n\times d_f}$ (as illustrated in Fig \ref{fig2}-(a)), where $n$ is the number of channels, $d_f$ is the dimension of the features, and $s_i$ is a $d_f$ dimensional vector that represents the time-frequency features of the EEG signal from the $i$th channel. To transform this sequence into a mesh grid, we use a transformation matrix $T\in\mathbb R^{m\times m}$, where $m$ represents the size of the matrix. The values in matrix $T$ are determined based on the spatial distribution of electrodes in the EEG device. Electrode locations are mapped onto the matrix by padding the matrix with electrode values in their corresponding positions and assigning zeros to positions without electrodes. In the case of our EEG system, the electrode sequence can be mapped into a $9\times 9$ matrix. The transformation matrix is defined as follows:
\begin{equation}
    \tiny{T=\begin{bmatrix}
        0 & 0 & FP1 & 0 & FPZ & 0 & FP2 & 0 & 0\\
        F9 & AF7 & AF3 & 0 & 0 & 0 & AF4 & AF8 & F10\\
        F7 & F5 & F3 & F1 & FZ & F2 & F4 & F6 & F8\\
        FT7 &  FC5  & FC3 & FC1 & FCZ & FC2 & FC4 & FC6 & FT8\\
        T7 & C5 & C3 & C1 & CZ & C2 & C4 & C6 & T8\\
        TP7 & CP5 & CP3 & CP1 & CPZ & CP2 & CP4 & CP6 & TP8\\
        P7 & P5 & P3 & P1 & PZ & P2 & P4 & P6 & P8\\
        P9 & PO7 & PO3 & 0 & POZ & 0 & PO4 & PO8 & P10\\
        0 & 0 & O1 & 0 & OZ & 0 & O2 & 0 & 0
    \end{bmatrix}}.
\end{equation}
Through the spatial mapping, the original EEG sequence ${[s_1, s_2, \cdots, s_n]}^T\in\mathbb R^{n\times d_f}$ can be transformed into a 3-D spatial representation like
\begin{equation}
    \begin{bmatrix}t_{11} & \cdots & t_{1m}\\ \vdots&\ddots&\vdots\\t_{m1} & \cdots & t_{mm}
\end{bmatrix}_{m \times m} ^T\in\mathbb R^{m^2\times d_f}.
\end{equation}
This spatial representation retains the $d_f$-dimensional feature information while also encoding the spatial arrangement of electrodes.

\subsection{Vision Transformer Model}
In our study, the extracted DE features with 3-D spatial representation could be considered as an image (EEGimage). Then we utilize the vision Transformer (ViT) architecture \cite{Dosovitskiy A}, a variant of Transformer, to introduce the spatial attention mechanism for the task of trust recognition from EEG data. Fig. \ref{fig2}-(b) provides an overview of our model. Since the standard Vision Transformer expects a sequence of token embeddings as input, we first reshape the 3-D EEGimage into a sequence of flattened 2-D fixed-size image patches before feeding it into the model. In particular, we divide the 3-D EEGimage $x \in\mathbb R^{m^2\times d_f}$ into sequences of patches $x_p \in\mathbb R^{L \times (p^2 \times d_f)}$, where $m^2=p^2 \times L$, $p$ represents the patch size, and $L$ is the number of patches. The Transformer model maintains a consistent latent vector size $D$ across all of its layers. Thus, we flatten the patches and map them to $D$ dimensions using a trainable linear projection $E\in \mathbb R ^{(p^2\times d_f)\times D}$.
Similar to how BERT employs a learnable embedding for tokens, we utilize a learnable embedding $z_0^0=x_{class}\in \mathbb R^{D}$ for the patch embeddings. We also include position embedding $E_{pos} \in \mathbb R^{(L+1)\times D}$to encode positional information. The resulting patch embeddings $z_0\in \mathbb R^{(L+1)\times D}$ are obtained as follows:
\begin{equation}
    z_0 = [x_{class};x_p^1E;x_p^2E;\cdots;x_p^LE]+E_{pos}.
\end{equation}
% \subsubsection{Transformer Architecture}

The embedded patches are then passed into the Vision Transformer encoders, which primarily consist of six Multiheaded Self-Attention (MSA) and MLP blocks. Layer normalization(Norm) and residual connections are applied before and following each block, respectively.
\begin{equation}
    z_{l}^{'} = MSA(LN(z_{l-1}))+z_{l-1}, (l\in[1,L]).
\end{equation}
\begin{equation}
    z_{l} = MLP(LN(z_{l}^{'}))+z_{l}^{'}, (l\in[1,L]).
\end{equation}
% This architectural design ensures the effective capture of spatial features within the EEG data using the ViT model.

\section{Result and Discussion}
In this section, we present our statistical analysis of the constructed dataset to evaluate the effectiveness of the varying trust level stimuli among different ability-level robots and task complexity. We then verify the trust recognition performance of our proposed model. Additionally, we have also conducted an ablation study to reveal the contribution of the spatial representation to the trust performance.
\subsection{Statistical Analysis of the Dataset}
\begin{figure}[!t]
    \centering
    \includegraphics[width=1.0\linewidth]{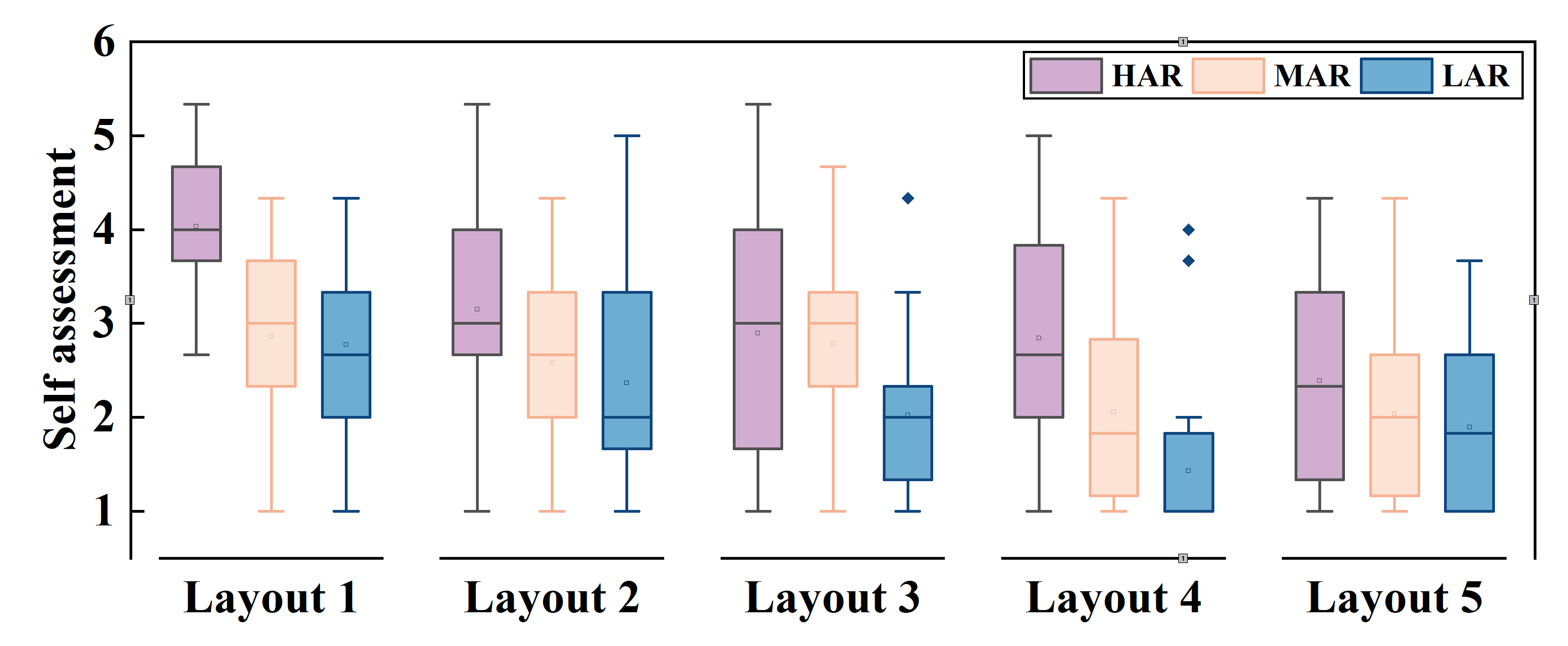}
    \caption{The distribution of participants' trust ratings in robots for different task difficulties and robot abilities. The rectangles are the $25\% \sim 75\%$ distribution of the self-assessment. The horizontal line at both ends is the range within 1.5IQR. The intermediate lines are the median line. The hollow points are the mean value. The solid points are the outliers.}
    \label{fig3}
\end{figure}

The subjective ratings provided by the participant regarding task difficulty and robot ability levels are reported in Fig. \ref{fig3}. For each layout, the trust levels vary depending on the robot's ability level. HAR generally has higher trust levels, while LAR has the lowest levels. MAR falls in between. Regarding task complexity, simpler tasks in Layouts 1, 2, and 3 result in higher overall trust scores and clear trust distinctions. However, in Layouts 4 and 5, where the scenarios were more complex, pre-trained robots struggled with the tasks, leading to lower overall trust scores.

In summary, there are significant statistical differences in trust ratings among different conditions, indicating successful elicitation of desired trust states. Besides, there are variances in trust ratings, indicating intra- and inter-individual variation. It is worth noting that trust states tend to have greater variance than distrust states. This can be easily explained by the fact that humans typically tend to have less doubt when not trusting something or someone, but more doubt when choosing to trust.

\subsection{Trust Recognition Performance}
\subsubsection{Experimental Setting}
We evaluate the trust recognition performance of the proposed model through two types of experiment settings: i) {\bf{\emph{slice-wise cross-validation}}} and ii) {\bf{\emph{trial-wise cross-validation}}}. In the first experiment setting, we randomly shuffle the slices among trials for each subject and then divide them into 10 folds. One of these folds is selected as testing data. The remaining nine folds are utilized as training data. The evaluation process is repeated 10 times, ensuring that each fold has been the testing data once. In the second experiment setting, we split the data from different trials for one subject into training and testing sets, following the methodology of \cite{Y. Ding}. In particular, 80\% of the trials are used for the training set, and the remaining 20\% trials as the testing set. The key difference between the two validation approaches lies in whether the slices within a single trial appear in both the training and testing datasets. 
Randomly shuffling the slices among trials before the training-testing split could result in slices within a single trial being present in both the training and testing data. However, in real-time scenarios, the highly correlated slices are never seen by the model.
On the contrary, trial-wise cross-validation makes sure that slices within a single trial will not appear in both the training and testing datasets. Therefore, the second experiment setting is more consistent with real applications and can further evaluate the model's generalization across trials.

To demonstrate the superior performance of our model, we compare the proposed model with traditional baseline models including Naive Bayesian (NB) \cite{Box G E P}, K-Nearest Neighbor (KNN) \cite{Cover T}, Support Vector Machine (SVM) \cite{Cortes C}, and one deep learning-based model, CNN. Various metrics are used to evaluate the performance of models to predict trust, including accuracy, F1 score, receiver operating characteristic curve (ROC), and the area under the ROC (AUC).
% Accuracy is commonly used for classifying tasks which measures the proportion of correctly classified samples. The F1 score, which is defined as the arithmetic average of precision and recall scores, is more effective for uneven data distribution. The AOC is a probability curve that describes the diagnostic ability of a classifier by illustrating the relationship between true positive rate (TPR) and false positive rate (FPR). The AUC, a quantitative indicator of AOC, represents the model's ability to distinguish between classes and measures the degree of separability.
The code is implemented using the PyTorch library. To ensure a fair comparison, all subjects were assigned the same hyperparameters. The batch size is set as 128, with a learning rate of 1e-4. The maximum training epoch is 500.  The training process was optimized using the Adam optimizer. To guide the training, cross-entropy loss is selected as the loss function. Both training and testing data were shuffled.

\subsubsection{Slice-wise Cross-validation Results}
\begin{figure}
    \centering
    \includegraphics[width=1.0\linewidth]{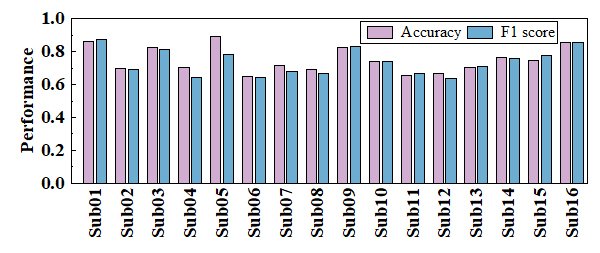}
    \caption{Mean accuracy and F1 score of each subject for trust recognition using our proposed model.}
    \label{fig4}
\end{figure}

The accuracy and F1 score for each subject using the proposed model are displayed in Fig. \ref{fig4}. 
The accuracies and F1 scores achieved by our model across all subjects are distributed from about 65\% to 90\%.
Notably, the model exhibits the best performance for Sub05, boasting an accuracy of 89.00\% and a F1 score of 78.59\%. Conversely, the accuracy and F1 score for Sub04 are the lowest, with 65.14\% and 64.16\%, respectively. 
These results reveal that our proposed model consistently demonstrates good trust recognition performance during human-robot cooperation using EEG for each subject. Nonetheless, the recognition performance still views individual differences.

To quantitatively compare the performance of the proposed method and baseline methods, the average accuracy and F1 score across all subjects are shown in Table \ref{table1}. The results reveal that our proposed model achieves the highest accuracy of 74.99\%. CNN achieves the highest F1 score of 73.79\%. Among the traditional models, SVM performs the best, with an accuracy of 70.82\% and an F1 score of 68.12\%. Our model surpasses the SVM model by 4.17\% in accuracy and 5.29\% in F1 score. These experimental results demonstrate that the two deep learning-based models have similar performance and significantly outperform traditional methods in the trial-wise cross-validation experiment.

\begin{table}[!t]
    \centering
    \caption{Slice-wise cross-validation results: mean and standard deviation of accuracy and F1 score of the proposed model and baseline models.}
    \setlength{\tabcolsep}{4mm}{
    \begin{tabular}[width=1.0\linewidth]{ccccc}
    \toprule 
    Method & ACC & std & F1 & std \\
    \hline
    NB & 58.98\% & 6.18\% & 54.69\% & 12.99\%\\
    KNN & 65.35\% & 6.50\% & 63.15\% & 6.07\%\\
    SVM & 70.82\% & 7.96\% & 68.12\% & 8.14\%\\ 
    CNN & 74.19\% & 6.34\% & \bf{73.79\%} & 7.07\%\\ 
    Our model & \bf{74.99\%} & 7.60\% & 73.41\% & 7.95\%\\
    \bottomrule     
    \end{tabular}}
    \label{table1}
\end{table}

\subsubsection{Trial-wise Cross-validation Results}
Table II displays the results of the trial-wise cross-validation. Our proposed method outperforms the others, achieving a 62.00\% accuracy and 62.59\% F1 score, respectively. In contrast, the CNN method exhibits the lowest performance, failing to recognize trust across trials. Among the traditional models, SVM performs the best. Notably, our model surpasses SVM, boasting a 0.10\% higher accuracy and a 4.19\% higher F1 score.

Comparing the results between slice-wise and trial-wise cross-validation, a general trend emerges where most models experience a decrease in the trial-wise cross-validation. Specifically, CNN fails to achieve cross-trial recognition, indicating a significant performance drop. In our method, accuracy decreases by 12.99\%, and F1 score decreases by 10.82\%. The performance gaps between the two experiment settings may be because trust is a continuous cognitive process in the brain, where slices within the same trials are highly homologous. In the slice-wise cross-validation, the slices from a single trial are included in both the training set and the test set. This setup allows models to capture the trial-specific features, potentially leading to higher performance. 

As for the generalization of different models, our model manifests promising performance and a reasonable level of generalization. However, it is noted that both deep learning-based methods exhibit more decreases in performance in the trial-wise cross-validation compared to the traditional methods. This suggests that most deep learning methods tend to have lower generalization compared to SVM and KNN. These findings emphasize the need for further research to enhance the generalization of our model in trust recognition tasks across trials.

\begin{table}[!t]
    \centering
    \caption{Trial-wise cross-validation results: mean and standard deviation of accuracy and F1 score of the proposed model and baseline models.}
    \setlength{\tabcolsep}{10.5pt}{
    \begin{tabular}[width=1.0\linewidth]{ccccc}
    \toprule 
    Method & ACC & std & F1 & std \\
    \hline
    NB & 58.47\% & 5.43\% & 57.11\% & 15.02\%\\
    KNN & 57.24\% & 8.55\% & 56.98\% & 9.74\%\\
    SVM & 61.90\% & 6.69\% & 58.40\% & 11.96\%\\
    CNN & 49.97\% & 7.55\%  & 41.38\% & 11.17\%\\
    Our model & \bf{62.00\%} & 7.96\% & \bf{62.59\% }& 7.66\%\\
    % ViT\_nosp & \bf{89.00\%} & 8.10\% & \bf{78.59\%} & 8.27\%\\
    \bottomrule     
    \end{tabular}}
    \label{table2}
\end{table}

\subsection{Ablation Study}
In this study, we propose the 3-D spatial representation for EEG and adapt the Vision Transformer to introduce spatial attention mechanisms to capture the topological information of the brain for the recognition of human trust levels. 
To further investigate the contribution of the spatial representation to the model performance, we have conducted an ablation experiment. This experiment involves comparing the performance of the proposed models with (Our model\_SP) and without spatial representation (Our model\_noSP), across both slice-wise and trial-wise cross-validation experiments.

\begin{figure}
    \centering
    \begin{subfigure}[b]{0.22\textwidth}
        \centering
        \includegraphics[width=\textwidth]{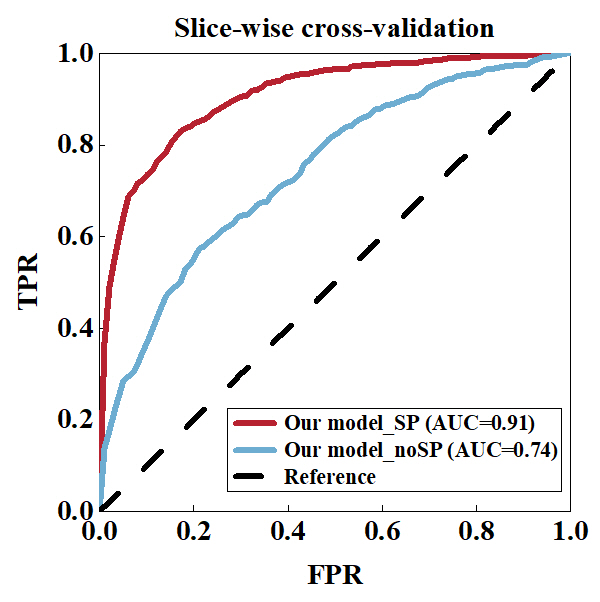}
        \caption{}
    \end{subfigure}
    \begin{subfigure}[b]{0.22\textwidth}
        \centering
        \includegraphics[width=\textwidth]{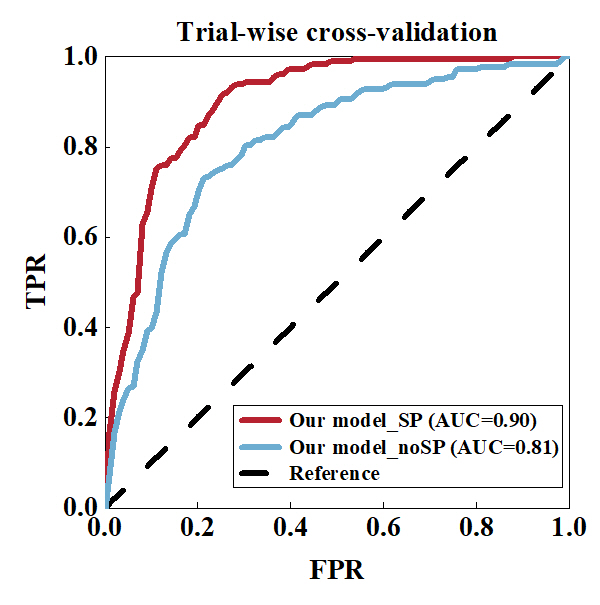}
        \caption{}
    \end{subfigure}
\caption{ROC curve and AUC of the EEG trust recognition models with or without spatial representation. (a) Slice-wise cross-validation. (b) Trial-wise cross-validation.}
\label{fig5}
\end{figure}

The results of the ablation study are summarized in Table III. Additionally, to further demonstrate the effectiveness of the spatial representation, we select Sub05 with the best recognition performance as a representative subject and depict the AOC and AUC in Fig. \ref{fig5}.
As can be seen, the ROC curves of models with spatial representation tend to be closer to the upper-left corner compared to those without spatial representation. Regarding average performance across subjects from Table III, our proposed approach with spatial representation achieves a 13.24\% higher accuracy and a 12.24\% higher in slice-wise cross-validation compared to the approach without spatial representation. Similarly, in trial-wise cross-validation, our method with spatial representation achieves a 6.92\% higher accuracy and a 12.76\% higher F1 score. These results reveal the effectiveness of our 3-D spatial representation in capturing spatial information of EEG data, leading to significant enhancements in recognizing human trust levels.

\begin{table}[!t]
    \centering
    \caption{Ablation study results: mean and standard deviation of accuracy and F1 score of our proposed model with or without spatial representation.}
    \setlength{\tabcolsep}{4.2pt}{
    \begin{tabular}[width=1.0\linewidth]{cccccc}
    \toprule 
    Experiment & Method & ACC & std & F1 & std \\
    \hline
    \multirow{2}{*}{Slice-wise}
    & Our model\_sp & \bf{74.99\%} & 7.60\% & \bf{73.41\%} & 7.95\%\\
    & Our model\_nosp & 61.75\% & 4.48\% & 61.17\% & 5.41\%\\
    \multirow{2}{*}{Trial-wise}
    & Our model\_sp & \bf{62.00\%} & 7.96\% & \bf{62.59\%} & 7.66\%\\
    & Our model\_nosp & 56.39\% & 6.52\% & 49.83\% & 13.22\%\\
    \bottomrule     
    \end{tabular}}
    \label{table2}
\end{table}
	
\section{CONCLUSIONS}
In this study, we achieved successful trust recognition in human-robot cooperation using EEG. We introduce a 3-D spatial representation and adapt the Vision Transformer for capturing spatial information. To validate our approach, we construct a public dataset, EEGTrust, for EEG-based human trust recognition during human-robot cooperation. Slice-wise and trial-wise cross-validation results demonstrate that our model surpasses baseline models in accuracy and generalization for deployment.

While our work shows promise, a limitation is the performance variability across subjects. Future efforts will concentrate on enhancing our model's adaptability and robustness to individual differences. Additionally, EEG-based trust measurement offers potential for real-time implementation, driving our focus towards developing a software architecture for facilitating trust recognition in real-time. 
Another aspect of our future work involves analyzing the relationship among EEG signals from different brain regions to enhance trust measurement precision and contribute to neuroscience advancements.


\newpage
\begin{thebibliography}{99}
\bibitem{Yuan L}
Yuan L, Gao X, Zheng Z, et al. In situ bidirectional human-robot value alignment[J]. Science robotics, 2022, 7(68): eabm4183.
\bibitem{Akash K1}
Akash K, McMahon G, Reid T, et al. Human Trust-based Feedback Control: Dynamically varying automation transparency
to optimize human-machine interactions[J]. IEEE Control Systems Magazine, 2020, 40(6): 98-116.
\bibitem{Xie Y}
Xie Y, Bodala I P, Ong D C, et al. Robot capability and intention in trust-based decisions across tasks[C]//2019 14th ACM/IEEE International Conference on Human-Robot Interaction (HRI). IEEE, 2019: 39-47.
\bibitem{Saeidi H}
Saeidi H, Wang Y. Incorporating trust and self-confidence analysis in the guidance and control of (semi) autonomous mobile robotic systems[J]. IEEE Robotics and Automation Letters, 2018, 4(2): 239-246.
\bibitem{Sadrfaridpour B}
Sadrfaridpour B, Wang Y. Collaborative assembly in hybrid manufacturing cells: an integrated framework for human–robot interaction[J]. IEEE Transactions on Automation Science and Engineering, 2017, 15(3): 1178-1192.
\bibitem{Xu A}
Xu A, Dudek G. Optimo: Online probabilistic trust inference model for asymmetric human-robot collaborations[C]//2015 10th ACM/IEEE International Conference on Human-Robot Interaction (HRI). IEEE, 2015: 221-228.
\bibitem{Chen M}
Chen M, Nikolaidis S, Soh H, et al. Trust-aware decision making for human-robot collaboration: Model learning and planning[J]. ACM Transactions on Human-Robot Interaction (THRI), 2020, 9(2): 1-23.
\bibitem{Malle B F}
Malle B F, Ullman D. A multidimensional conception and measure of human-robot trust[M]//Trust in Human-Robot Interaction. Academic Press, 2021: 3-25.
\bibitem{Jian J Y}
Jian J Y, Bisantz A M, Drury C G. Foundations for an empirically determined scale of trust in automated systems[J]. International journal of cognitive ergonomics, 2000, 4(1): 53-71.
\bibitem{Merritt S M}
Merritt S M. Affective processes in human–automation interactions[J]. Human Factors, 2011, 53(4): 356-370.
\bibitem{J. Y. C. Chen}
J. Y. C. Chen and P. I. Terrence, “Effects of imperfect automation and individual differences on concurrent performance of military and robotics tasks in a simulated multitasking environment,” Ergonomics, vol. 52, no. 8, pp. 907–920, 2009.
\bibitem{Baker} C. L. Baker, R. Saxe, and J. B. Tenenbaum, “Action understanding as inverse planning,” Cognition, vol. 113, no. 3, pp. 329–349, 2009. doi:10.1016/j.cognition.2009.07.005.

\bibitem{Freedy A}
Freedy A, DeVisser E, Weltman G, et al. Measurement of trust in human-robot collaboration[C]//2007 International
Symposium on Collaborative Technologies and Systems. IEEE, 2007: 106-114.
\bibitem{Miller D}
Miller D, Johns M, Mok B, et al. Behavioral measurement of trust in automation: the trust fall[C]//Proceedings of the human factors and ergonomics society annual meeting. Sage CA: Los Angeles, CA: SAGE Publications, 2016, 60(1): 1849-1853.
\bibitem{Wright T J}
Wright T J, Horrey W J, Lesch M F, et al. Drivers’trust in an autonomous system: Exploring a covert video-based measure
of trust[C]//Proceedings of the Human Factors and Ergonomics Society Annual Meeting. Sage CA: Los Angeles, CA: SAGE
Publications, 2016, 60(1): 1334-1338.
\bibitem{Lu Y}
Lu Y, Sarter N. Eye tracking: a process-oriented method for inferring trust in automation as a function of priming and
system reliability[J]. IEEE Transactions on Human-Machine Systems, 2019, 49(6): 560-568.
\bibitem{Aimone}
Aimone J A, Houser D, Weber B. Neural signatures of betrayal aversion: an fMRI study of trust[J]. Proceedings of the Royal Society B: Biological Sciences, 2014, 281(1782): 20132127.
\bibitem{Baumgartner}
Baumgartner T, Heinrichs M, Vonlanthen A, et al. Oxytocin shapes the neural circuitry of trust and trust adaptation in humans[J]. Neuron, 2008, 58(4): 639-650.
\bibitem{Riedl}
R. Riedl and A. Javor, “The biology of trust: Integrating evidence from genetics, endocrinology, and functional brain imaging,” J. Neurosci. Psychol. Econ., vol. 5, no. 2, pp. 63–91, 2012.
\bibitem{Oh S}
Oh S, Seong Y, Yi S, et al. Neurological measurement of human trust in automation using electroencephalogram[J]. International Journal of Fuzzy Logic and Intelligent Systems, 2020, 20(4): 261-271.
\bibitem{Wang M}
Wang M, Hussein A, Rojas R F, et al. EEG-based neural correlates of trust in human-robot interaction[C]//2018 IEEE Symposium Series on Computational Intelligence (SSCI). IEEE, 2018: 350-357.
\bibitem{Akash K}
Akash K, Hu W L, Jain N, et al. A classification model for sensing human trust in machines using EEG and GSR[J]. ACM Transactions on Interactive Intelligent Systems (TiiS), 2018, 8(4): 1-20.
\bibitem{Choo}
Choo S, Nam C S. Detecting human trust calibration in automation: a convolutional neural network approach[J]. IEEE Transactions on Human-Machine Systems, 2022, 52(4): 774-783.
\bibitem{Ajenaghughrure}
Ajenaghughrure I B, Sousa S C D C, Lamas D. Psychophysiological Modeling of Trust In Technology: Influence of Feature Selection Methods[J]. Proceedings of the ACM on Human-Computer Interaction, 2021, 5(EICS): 1-25.
\bibitem{Bassett}
D. S. Bassett and O. Sporns, “Network neuroscience,” Nature Neurosci., 650 vol. 20, no. 3, pp. 353–364, Feb. 2017.
\bibitem{Carroll}
Carroll, Micah, Rohin Shah, Mark K. Ho, Thomas L. Griffiths, Sanjit A. Seshia, Pieter Abbeel, and Anca Dragan. "On the utility of learning about humans for human-ai coordination." NeurIPS 2019.
\bibitem{Hancock} P. A. Hancock, T. T. Kessler, A. D. Kaplan, J. C. Brill, and J. L. Szalma, “Evolving Trust in Robots: Specification through sequential and comparative meta-analyses,” Human Factors: The Journal of the Human Factors and Ergonomics Society, vol. 63, no. 7, pp. 1196–1229, 2020.
\bibitem{Hyvärinen}
A. Hyvärinen and E. Oja, “Independent component analysis: Algorithms and applications,” Neural Netw., vol. 13, nos. 4–5, pp. 411–430, Jun. 2000.
\bibitem{Delorme}
A. Delorme and S. Makeig, “EEGLAB: An open source toolbox for analysis of single-trial EEG dynamics including independent component analysis,” J. Neurosci. Methods, vol. 134, no. 1, pp. 9–21, 2004.
\bibitem{Schirrmeister}
R. T. Schirrmeister, J. T. Springenberg, L. D. J. Fiederer, M. Glasstetter, K. Eggensperger, M. Tangermann, F. Hutter, W. Burgard, and T. Ball, “Deep learning with convolutional neural networks for EEG decoding and visualization,” Human Brain Mapping, vol. 38, no. 11, pp. 5391–5420, 2017.
\bibitem{R.-N. Duan}
R.-N. Duan, J.-Y. Zhu, and B.-L. Lu, “Differential entropy feature for EEG-based emotion classification,” in 6th International IEEE/EMBS Conference on Neural Engineering (NER), IEEE, 2013, pp. 81–84.
\bibitem{Dosovitskiy A}
Dosovitskiy A, Beyer L, Kolesnikov A, et al. An Image is Worth 16x16 Words: Transformers for Image Recognition at Scale[C]//International Conference on Learning Representations.
\bibitem{Y. Ding}
Y. Ding, N. Robinson, S. Zhang, Q. Zeng and C. Guan, "TSception: Capturing Temporal Dynamics and Spatial Asymmetry from EEG for Emotion Recognition," in IEEE Transactions on Affective Computing, doi: 10.1109/TAFFC.2022.3169001.
\bibitem{Box G E P}
Box G E P, Tiao G C. Bayesian inference in statistical analysis[M]. John Wiley \& Sons, 2011.
\bibitem{Cover T}
Cover T, Hart P. Nearest neighbor pattern classification[J]. IEEE transactions on information theory, 1967, 13(1): 21-27.
\bibitem{Cortes C}
Cortes C, Vapnik V. Support-vector networks[J]. Machine learning, 1995, 20: 273-297.
% \bibitem{c1} G. O. Young, ÒSynthetic structure of industrial plastics (Book style with paper title and editor),Ó 	in Plastics, 2nd ed. vol. 3, J. Peters, Ed.  New York: McGraw-Hill, 1964, pp. 15Ð64.
% \bibitem{c2} W.-K. Chen, Linear Networks and Systems (Book style).	Belmont, CA: Wadsworth, 1993, pp. 123Ð135.
% \bibitem{c3} H. Poor, An Introduction to Signal Detection and Estimation.   New York: Springer-Verlag, 1985, ch. 4.
% \bibitem{c4} B. Smith, ÒAn approach to graphs of linear forms (Unpublished work style),Ó unpublished.
% \bibitem{c5} E. H. Miller, ÒA note on reflector arrays (Periodical styleÑAccepted for publication),Ó IEEE Trans. Antennas Propagat., to be publised.
% \bibitem{c6} J. Wang, ÒFundamentals of erbium-doped fiber amplifiers arrays (Periodical styleÑSubmitted for publication),Ó IEEE J. Quantum Electron., submitted for publication.
% \bibitem{c7} C. J. Kaufman, Rocky Mountain Research Lab., Boulder, CO, private communication, May 1995.
% \bibitem{c8} Y. Yorozu, M. Hirano, K. Oka, and Y. Tagawa, ÒElectron spectroscopy studies on magneto-optical media and plastic substrate interfaces(Translation Journals style),Ó IEEE Transl. J. Magn.Jpn., vol. 2, Aug. 1987, pp. 740Ð741 [Dig. 9th Annu. Conf. Magnetics Japan, 1982, p. 301].
% \bibitem{c9} M. Young, The Techincal Writers Handbook.  Mill Valley, CA: University Science, 1989.
% \bibitem{c10} J. U. Duncombe, ÒInfrared navigationÑPart I: An assessment of feasibility (Periodical style),Ó IEEE Trans. Electron Devices, vol. ED-11, pp. 34Ð39, Jan. 1959.
% \bibitem{c11} S. Chen, B. Mulgrew, and P. M. Grant, ÒA clustering technique for digital communications channel equalization using radial basis function networks,Ó IEEE Trans. Neural Networks, vol. 4, pp. 570Ð578, July 1993.
% \bibitem{c12} R. W. Lucky, ÒAutomatic equalization for digital communication,Ó Bell Syst. Tech. J., vol. 44, no. 4, pp. 547Ð588, Apr. 1965.
% \bibitem{c13} S. P. Bingulac, ÒOn the compatibility of adaptive controllers (Published Conference Proceedings style),Ó in Proc. 4th Annu. Allerton Conf. Circuits and Systems Theory, New York, 1994, pp. 8Ð16.
% \bibitem{c14} G. R. Faulhaber, ÒDesign of service systems with priority reservation,Ó in Conf. Rec. 1995 IEEE Int. Conf. Communications, pp. 3Ð8.
% \bibitem{c15} W. D. Doyle, ÒMagnetization reversal in films with biaxial anisotropy,Ó in 1987 Proc. INTERMAG Conf., pp. 2.2-1Ð2.2-6.
% \bibitem{c16} G. W. Juette and L. E. Zeffanella, ÒRadio noise currents n short sections on bundle conductors (Presented Conference Paper style),Ó presented at the IEEE Summer power Meeting, Dallas, TX, June 22Ð27, 1990, Paper 90 SM 690-0 PWRS.
% \bibitem{c17} J. G. Kreifeldt, ÒAn analysis of surface-detected EMG as an amplitude-modulated noise,Ó presented at the 1989 Int. Conf. Medicine and Biological Engineering, Chicago, IL.
% \bibitem{c18} J. Williams, ÒNarrow-band analyzer (Thesis or Dissertation style),Ó Ph.D. dissertation, Dept. Elect. Eng., Harvard Univ., Cambridge, MA, 1993. 
% \bibitem{c19} N. Kawasaki, ÒParametric study of thermal and chemical nonequilibrium nozzle flow,Ó M.S. thesis, Dept. Electron. Eng., Osaka Univ., Osaka, Japan, 1993.
% \bibitem{c20} J. P. Wilkinson, ÒNonlinear resonant circuit devices (Patent style),Ó U.S. Patent 3 624 12, July 16, 1990. 






\end{thebibliography}
\end{document}